\documentclass[aps,prl,twocolumn,superscriptaddress,amsfont,amssymb,amsmath,nofootinbib,showpacs,balancelastpage]{revtex4-1}
\usepackage{graphicx,longtable,mathrsfs,color}
\usepackage[usenames,dvipsnames]{xcolor} 
\usepackage{amssymb,amsmath,mathtools,mathrsfs} 
\usepackage{epsfig,subfigure} 
\usepackage{booktabs,longtable} 
\usepackage{exscale,relsize} 
\usepackage[normalem]{ulem} 
\usepackage{enumerate} 
\usepackage{hyperref}

\makeindex

\begin{document}

\title{Noether Identities and Gauge-Fixing the Action for Cosmological Perturbations}

\author{Macarena Lagos}
\email{m.lagos-urbina13@imperial.ac.uk}
\affiliation{Departamento de F\'{\i}sica, P. Universidad Cat\'{o}lica de Chile, Casilla 306, Santiago, Chile.}
\affiliation{Theoretical Physics, Blackett Laboratory, Imperial College, Prince Consort Road, London SW7 2BZ, UK}
\affiliation{Astrophysics, University of Oxford, DWB, Keble Road, Oxford OX1 3RH, UK}
\author{M\'{a}ximo Ba\~{n}ados}
\email{maxbanados@fis.puc.cl}
\affiliation{Departamento de F\'{\i}sica, P. Universidad Cat\'{o}lica de Chile, Casilla 306, Santiago, Chile.}
\author{Pedro G. Ferreira}
\email{p.ferreira1@physics.ox.ac.uk}
\affiliation{Astrophysics, University of Oxford, DWB, Keble Road, Oxford OX1 3RH, UK}
\author{Sebasti\'an Garc\'{\i}a-S\'aenz}
\email{sg2947@columbia.edu}
\affiliation{Department of Physics, Columbia University, New York, NY 10027, USA}
\date{\today}

\begin{abstract}
We propose and develop a general algorithm for finding the action for cosmological perturbations which rivals the conventional, gauge-invariant approach and can be applied to theories with more than one metric. We then apply it to a particular case of bigravity, focusing on the Eddington-inspired Born-Infeld theory, and show that we can obtain a nearly scale-invariant power spectrum for both scalar and tensor primordial quantum perturbations. Unfortunately, in the case of the minimal Eddington-inspired Born-Infeld theory, we find that the tensor-to-scalar ratio of perturbations is unacceptably large. We discuss the applicability of our general method  and the possibility of resurrecting the specific theory we have looked at.
\end{abstract}

\keywords{These are the keywords.}

\maketitle

\section{Introduction}

We would love to understand how the universe began. Due to the resounding successes of observational cosmology, we now have a tentative idea of what the very early universe was like.
It was probably smooth and hot yet ever so slightly perturbed, with ripples that can, with tremendous accuracy, be described as adiabatic, quasi scale-invariant and Gaussian. The recent results from the Planck experiment \cite{Ade:2013uln} have characterised these properties with exquisite precision and we are now confident that we can assume them in the subsequent formation of large scale structure. 

If we are to access the very beginning of time we need to extrapolate and to do so, we use General Relativity (GR), our most successful theory of gravity. Such an extrapolation is not exempt from problems for, if embraced wholeheartedly, it predicts the Big Bang, a physical divergence in which physical quantities such as the energy density diverge, and a horizon structure that is difficult to reconcile with one of our key assumptions - smoothness. The immensely successful theory of cosmological inflation is often invoked as naturally leading to a satisfactory explanation for the initial state of the universe. Yet, in detail, it has many problems. The simplest models fail to fit the observations and, more fundamentally, they rely on an incredibly fine tuned set of conditions to be viable. 

It pays to consider alternatives to the initial state of the universe. One intriguing arena to explore is that of multigravity, i.e. where more than one metric is at play. Particular examples of theories with multiple metrics have recently been looked at in detail when constructing consistent theories of massive gravitons \cite{deRham:2010kj,Hassan:2011zd}. An example that attempts to do away with the initial singularity is the Eddington-inspired Born-Infeld (EiBI) theory \cite{Vollick:2003qp,Banados:2010ix}. Originally motivated by Born-Infeld electrodynamics and Eddington's gravity, EiBI is a classical gravitational theory which introduces modifications to General Relativity in regions with large curvature. The theory can be formulated  \cite{Delsate:2012ky} as a bimetric-like theory:
\begin{align}\label{EBI}
S&[g,q,\chi] = -{1 \over 2} \int d^4x\;  \sqrt{q} \left(  R(q) + {2 \over \kappa}  \right) \\
&+{1 \over 2} \int d^4x {1 \over \kappa}(\sqrt{q}q^{\mu\nu}g_{\mu\nu} - 2\sqrt{g}   )+ S_{\text{m}}[\chi,g]   , \nonumber
\end{align}
where $q$ and $g$ are metrics with signature (+, -, -, -), $S_{\text{m}}$ is the matter action depending on a matter field $\chi$, and $\kappa$ is an arbitrary constant. Throughout this paper we will use $8\pi G=c=1$. Notice here that the metric representing the physical spacetime is only $g$ because that is the one coupled to matter. For our purpose, eq.~(\ref{EBI}) turns out to be a formulation clearer and simpler than that shown in \cite{Banados:2010ix}, so it will be used throughout this paper. It is an
exotic form of bigravity in which there is no kinetic term for the $g$ metric (which couples to matter) unlike in the case of massive gravity \cite{deRham:2010kj} where it is the $q$ metric which has no dynamics. Note that although in both theories there is a metric with no dynamics, these are actually very different. In the EiBI theory, the metric $g$ is an {\it auxiliary} field and hence is varied in the action. In massive gravity  \cite{deRham:2010kj}, the non-dynamical metric is a fixed, {\it reference} field and is therefore not varied in the action. This difference leads to more equations or constraints and fewer degrees of freedom in the EiBI theory than in the theory proposed in \cite{deRham:2010kj}.

In \cite{Banados:2010ix} it was shown that, in the context of cosmology, the EiBI theory avoids the Big Bang singularity by predicting a universe with a nearly static past. This type of evolution is  interesting because it eliminates a physical divergence but also generates an early stage of inflation without including any unknown type of matter. Subsequent authors have looked at various aspects of this theory \cite{Pani:2011mg, Pani:2012qd,Delsate:2012ky,Scargill:2012kg,EscamillaRivera:2012vz,Delsate:2013bt,Harko:2013xma,Cho:2013pea,Harko:2013wka,Harko:2013aya,Yang:2013hsa,Avelino:2012qe,Sham:2012qi,Cho:2012vg,EscamillaRivera:2013hv,Bouhmadi-Lopez:2013lha,Sham:2013sya,Jana:2013fga,Cho:2013usa,Avelino:2012ue,Liu:2012rc,Avelino:2012ge,Casanellas:2011kf}.

If we are to fully explore the primordial properties of the EiBI theory, and multigravity theories more generally, we need to develop a systematic and straightforward approach for studying linear perturbation theory around a homogeneous background. The origin of these perturbations via some putative quantum mechanism and how they later seed the formation of structure in the Universe are crucial in making them credible, testable theories. The introduction of extra tensor fields (as opposed to scalar fields as is usually done in inflation) greatly complicates the endeavour. There are multiple fields and gauge degrees of freedom which make it cumbersome to identify the true, gauge invariant physical fields. 

In this paper we develop a procedure for extracting the gauge-invariant physical fields from an initial set of perturbations (which are generally gauge-dependent) and finding an action for them. This method provides a general yet simple alternative to what has become known as the {\it Mukhanov variable} \cite{Mukhanov:1990me}  and relies on correctly identifying the various Noether identities and making what we call ``good" gauge choices. Given its generality, the method can be easily rolled out for the suite of bigravity theories that are currently being considered and we take great care in describing and explaining the different steps in some detail.

The internal working of EiBI is remarkably fruitful for developing these methods and we use them to predict the power spectra of scalar and tensor inhomogeneities in our universe by assuming, as usual, that they were generated by primordial quantum first-order perturbations in a classical cosmological background. We find that, indeed, we obtain scale invariant perturbations but to do so we invoke an unconventional mechanism proposed by Hollands and Wald \cite{Hollands:2002yb}. We find that the main drawbacks of the EiBI theory become manifest: tensor and scalar instabilities occur, where these type of perturbations diverge asymptotically in the past and we predict an unacceptably large fraction of tensor to scalar perturbations.

The main focus of this paper is the method and we explain it in some detail. After briefly describing the EiBI theory we lay out our method by, first of all, using the general relativistic case as an example. We then turn to a particular case of bigravity and apply our method to EiBI and find the fundamental scalar and tensor variables. We then quantise these variables and calculate the fundamental observables: the relative amplitude between the scalar and tensor perturbations as well as their scalar and tensor spectral indices. Finally we discuss our results.

\section{E\lowercase{i}BI Theory}
\label{EiBI}

The action for the EiBI theory was originally written in \cite{Banados:2010ix} as:
\begin{align}\label{EiBI2}
 S&[g,\Gamma, \chi]=\frac{1}{\kappa}\int d^4x \left[\sqrt{|g_{\mu\nu}-\kappa R_{\mu\nu}(\Gamma)|}-\lambda\sqrt{-g}\right]\nonumber \\
& + S_\text{m}[\chi,g],
\end{align}
where $g$ is the metric, $S_m$ is the matter action which depends on a matter field $\chi$, $\Gamma$ is an affine connection (independent of $g$), $\kappa$ an arbitrary constant with units of $L^2$, and $\lambda$ is a non-zero constant related to the cosmological constant. This action was constructed initially to have the same main characteristic as the Born-Infeld electrodynamic theory: eliminate divergences. As we will describe next, the EiBI theory actually avoids the Big Bang.

\begin{figure}[h]
\includegraphics[width=0.47\textwidth]{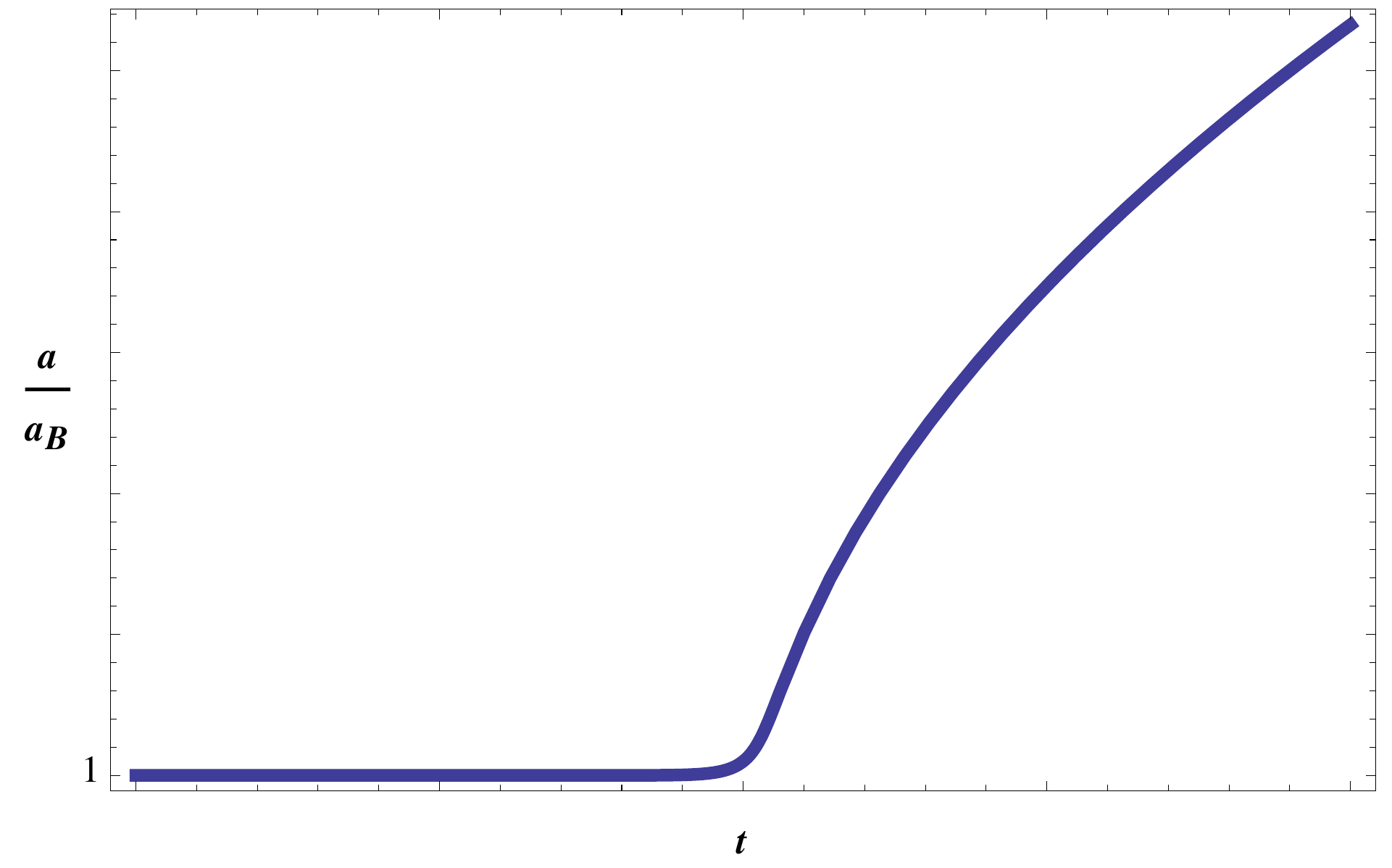}
\caption{Scale factor as a function of physical time $t$ during the radiation-dominated era. An asymptotic minimum value $a_B$ is found as $t\rightarrow -\infty$, which avoids the Big Bang divergence. During the Eddington regime the scale factor grows exponentially fast, while during the Einstein regime it evolves as predicted by GR.}
\label{fig1}
\end{figure}

Action (\ref{EiBI2}), with $\lambda=1$ (no cosmological constant), is completely equivalent to (\ref{EBI}) if one considers that $\Gamma$ in (\ref{EiBI2}) is the affine connection of the metric $q$ in (\ref{EBI}). From (\ref{EBI}), we find the following equations of motion:
\begin{align}\label{eq2}
& q_{\mu\nu}=g_{\mu\nu}-\kappa R_{\mu\nu}(q),\nonumber \\
 &\sqrt{-q}q^{\mu\nu}=\sqrt{-g}g^{\mu\nu}+\kappa T^{\mu\nu}, 
\end{align}

\noindent where $T^{\mu\nu}$ is the standard energy-momentum tensor (with indices raised with the metric $g$), satisfying a conservation equation: $T^{\mu\nu}{}_{;\nu}=0$, where the covariant derivative is with respect to $g$. For a flat FRW universe, both metrics are described by:
\begin{align}
 ds_q^2 &= b^2[z^{-1}d\eta^2 -d\vec{x} \cdot d\vec{x}],\nonumber\\
 ds_g^2 &= a^2[d\eta^2-d\vec{x} \cdot d\vec{x}], 
\end{align}
where $b$, $a$ and $z$ are all functions of the conformal time $\eta$, and parametrize the evolution of the expanding background. Since both metrics are coupled, these three parameters are related to each other, and then both metrics cannot be written in the FRW form at the same time.
In the case of a perfect fluid with an equation of state between the pressure $p$ and the rest energy density $\rho$ given by $p=w \rho$ (with $w$ constant), the Friedmann equation is the following:
\begin{align}\label{Fried}
& H^2= \frac{3}{\kappa}  (1+\kappa \rho)(1-\kappa \rho w)^2 \nonumber\\
& \times \frac{ \left[ \frac{1}{2} (1+3w)\kappa  \rho -1\right]+\sqrt{ (1+\kappa \rho) (1-\kappa \rho w)^3} }  {\left[3 + \frac{3}{2}w (1+3w)\kappa^2 \rho^2 + \frac{3}{4}(3w-1)(w-1)\kappa \rho\right]^2},
\end{align}
where $H=(da/dt)/a$ is the Hubble parameter, with $a$ the scale factor, and $t$ the physical time.
For small densities, i.e.~$\kappa\rho \ll 1$, this equation is equivalent to the Friedman limit of GR,
\begin{equation}
 H^2 =  \frac{\rho}{3}, \nonumber
\end{equation}
as we expected, since the EiBI theory modifies GR only for large curvatures.
The solution of the scale factor $a(t)$, satisfying eq.~(\ref{Fried}), is described in \cite{Banados:2010ix,Scargill:2012kg}. Two different types of behaviour are observed for $a(t)$, depending on the sign of $\kappa$, but we will focus only on the case $\kappa>0$, where the universe presents a minimum scale factor $a_B$ (whose value depends on $\kappa$) in the asymptotic past, i.e.~$t\rightarrow -\infty$. A scheme of such an evolution is shown in Fig.~\ref{fig1} for the radiation-dominated era ($w=1/3$). In this figure we can see that the scale factor initially stays near to the minimum (a stage we call the {\it Eddington} regime) and subsequently it evolves as predicted by GR (a stage we call the {\it Einstein} regime). 

During the Eddington regime, the scale factor evolves as:
\begin{equation}
  a(t)=a_B \left(e^{ \sqrt{\frac{8}{3\kappa}}(t-t_0)}+1\right),   \label{aPositivo} 
\end{equation}
which means that the universe undergoes an early accelerated exponential expansion away from the minimum scale factor, and therefore there is an inflationary period. This type of evolution is very interesting because it solves the horizon problem (one of the attractive features of  the theory of inflation in GR) without considering any unknown type of matter field. 
The horizon problem is avoided in this solution because the horizon gets infinitely large as we approach $a_B$. In particular, in the early universe, there was a time when the horizon was bigger than the size of the horizon at the time of the photon decoupling. The flatness problem is also alleviated (although not completely solved) in this model- during the Eddington regime the scale factor does not grow much, and $\Omega$ is roughly constant. It is only during the Einstein regime that $\Omega$ starts to deviate from unity. As a result, the EiBI theory gives us a well behaved background that suggests a possible alternative to inflation.

The main prediction from the theory of inflation is a well-defined set of initial conditions for structure formation. We will attempt to do the same for EiBI. As in inflation, we will assume that primordial quantum perturbations were originated during the early universe, i.e.~the radiation-dominated era, transforming to classical perturbations which, through gravitational instability, grew until structures were formed. If we are to mimic what has been done for the theory of inflation, we need to consider the following steps: find the first-order perturbations for all the fields in (\ref{EBI}), calculate a second-order action for them, identify the second-order action for the {\it physical} fields only, and finally quantise these physical fields. In doing so, we will find their power spectrum which we can (almost) directly link to observations.

To calculate a second-order action for the gauge-invariant physical fields we will take an alternative route to the conventional approach of \cite{Mukhanov:1990me}. We will take advantage of the gauge symmetry present in the theory to eliminate redundant degrees of freedom and pin down the fundamental fields that we need to quantise. In order to make this process clear, in the next section we will calculate the second-order action for the scalar field that drives inflation in GR, reproducing the well-known result.

\section{The second-order action for GR}
\label{Method} 
In General Relativity, the simplest model of inflation involves a scalar field which satisfies the following action (\cite{Dodelson:2003, GarciaBellido:2005df}):
\begin{align} \label{GRaction}
 & S[g,\varphi] = -\frac{1}{2} \int d^4x\; \sqrt{-g}R\nonumber\\
&+\int d^4x\; \sqrt{-g}\left( \frac{1}{2}\partial^\mu \varphi \partial_\mu\varphi -V(\varphi) \right),
\end{align}
which has a gauge symmetry that comes from considering general coordinate transformations. We consider general first-order perturbations for the metric $g$ and the inflationary scalar field $\varphi$, in a background given by a flat FRW universe with a scalar $\varphi_0(\eta)$. Following the classification of \cite{Mukhanov:1990me} in scalar, vector and tensor perturbations (which evolve independently), we consider only, as an example, the scalar modes:
\begin{align} \label{eq3}
 ds_g^2 &= a^2[(1+2\phi)d\eta^2-2 B_{,i}dx^id\eta \nonumber\\
&-[(1-2\psi)\delta_{il}+2 E_{,il}]dx^i dx^l],\nonumber \\
 \varphi&=\varphi_0+\varphi_1,
\end{align} 
where the first-order perturbations are represented by the fields $\phi$, $B$, $\psi$, $E$ and $\varphi_1$, which all depend, in general, on $\vec{x}$ and $\eta$. If we replace (\ref{eq3}) into (\ref{GRaction}) and Taylor expand up to second order in the perturbation fields, we find\footnote{In principle, one would have expected that some second-order perturbation fields (that were not considered in our calculations) contributed also to the second-order action, but they do not as these terms would appear multiplying the background equations of motion that hold. This fact is a generic property of perturbed actions.}:
\begin{equation}\label{2action}
 S_\text{s}[\phi, B, \psi, E, \varphi_1] = S_{\text{gs}}+S_{\text{ms}},
\end{equation}
where $S_{\text{gs}}$ corresponds to the first integral of (\ref{GRaction}), i.e.~the purely gravitational part, and $S_{\text{ms}}$ corresponds to the second integral of (\ref{GRaction}), i.e.~the matter action. Explicitly, 
\begin{align}
&S_{\text{gs}}[\phi, B, \psi, E]=\frac{1}{2}\int d^4x\; a^2 \left[-6\psi^{'2} -12\mathcal{H}(\phi+\psi)\psi'\frac{}{}\right.\nonumber \\
&-9\mathcal{H}^2(\phi+\psi)^2-2\psi_{,i}(2\phi_{,i}-\psi_{,i} )+4\mathcal{H}\psi'E_{,ii}  \nonumber \\
& +3\mathcal{H}^2B_{,i}B_{,i}-4\mathcal{H}(\phi+\psi)(B-E')_{,ii} -4\psi' (B-E')_{,ii} \nonumber\\
& -4\mathcal{H}\psi_{,i}B_{,i}+6\mathcal{H}^2(\psi+\phi)E_{,ii}-4\mathcal{H}E_{,ii}(B-E')_{,ll}\nonumber\\
&\left.+4\mathcal{H}E_{,ii}B_{,ll}+3\mathcal{H}^2E_{,ii}E_{,ll}\right]
\end{align}
\begin{align}
&S_{\text{ms}}[\phi, B, \psi, E, \varphi_1] = \frac{1}{2}\int d^4x\; a^2 \left[ \varphi_0^{'2}\left(4\phi^2-B_{,i}B_{,i}\right)\right.\nonumber\\
&+\left(\varphi_0^{'2} -2 V_0 a^2  \right)\left(\frac{1}{2}(3\psi^2-\phi^2+B_{,i}B_{,i}-E_{,ii}E_{,ll})\right. \nonumber\\
&\left.-3\phi\psi+(\phi-\psi) E_{,ii}\frac{}{}\right)-2\varphi_0'\varphi_{1,i}B_{,i}-4\varphi_0'\varphi_1'\phi+\varphi_1^{'2}\nonumber\\
&+ 2\left(\phi-3\psi+E_{,ii}\right)\left(\varphi_1'\varphi_0'-V_1a^2-\phi\varphi_0^{'2}\right)-\varphi_{1,i}\varphi_{1,i}\nonumber\\
&\left.-2V_2a^2\right]
\end{align}
where we have represented the time derivatives of $\eta$ as $'$, and the spatial derivatives as ``$,i$" (for $i=1,2,3$), we have defined $\mathcal{H}\equiv a'/a$, and the potential $V$ has been expanded as $V=V_0(\varphi_0)+V_1(\varphi_1)+V_2(\varphi_1)$, such that $V_0$, $V_1$ and $V_2$ are of zeroth, first and second order, respectively. This second-order action gives first-order equations of motion and is invariant only under infinitesimal transformations of first order. This means, there are unphysical degrees of freedom, related to this gauge symmetry. Note that, in general, there could be yet more unphysical degrees of freedom coming from auxiliary fields with no dynamics (no time derivatives). 

In order to eliminate all the unphysical degrees we will fix the gauge in action (\ref{2action}) and eliminate the auxiliary fields. The elimination of auxiliary fields can be easily done by calculating their equations of motion (which are algebraic in the auxiliary fields) and working these fields out in terms of the fields with dynamics. However, fixing the gauge inside an action is a subtle process. The idea is to fix some functions (equaling to the number of gauge symmetries) to prescribed values, typically zero, in the action at the very beginning. Once these functions are set to zero, their corresponding equations of motion are lost and so is their information. 

A good gauge condition is one such that the equations of motion we get rid of are contained in the remaining ones, i.e.~are redundant equations. In other words, the gauge-fixed action must carry the same information as the original one (a simple example that makes this statement clear can be found in the Appendix). In addition, a good gauge condition must fix the gauge. This means that after some functions are set to zero, there is no residual gauge freedom.
It turns out that the very existence of gauge invariance provides a simple and direct criteria to distinguish between good and bad gauge conditions. Let us briefly describe here the main ideas. 

The crucial ingredients are the Noether identities. These identities are relations between the various equations of motion, and follow directly from the gauge invariance of the action. Action (\ref{2action}) is invariant under (see \cite{Mukhanov:1990me}):
\begin{align}\label{GaugeTransf1}
 &\delta\phi=-(a'/a)\xi^{0}-\xi^{0'},\; \delta\psi=(a'/a)\xi^0, \nonumber \\
 & \delta B=\xi^0-\xi',\; \delta E=-\xi, \;\delta \varphi_1=-\varphi'_0\xi^{0},
\end{align}
where $\xi^{0}$ and $\xi$ are arbitrary first-order scalar functions, representing gauge freedoms. To find the Noether identities, we consider a general variation of the perturbed action (\ref{2action}):
\begin{equation}
\delta S_\text{s} = \int d^4x\; \left( {\cal E}_{\phi} \delta \phi  + {\cal E}_{B} \delta B + {\cal E}_{\psi} \delta \psi + {\cal E}_{E} \delta E + {\cal E}_{\varphi_1} \delta\varphi_1\right), 
\end{equation}

\noindent where ${\cal E}_{n}$ denotes the equation of motion for a field $n$. Now, if we replace these variations of fields $\delta n$ by the gauge variations (\ref{GaugeTransf1}), we must obtain $\delta S=0$, because the action is gauge invariant. Let us do this and perform some integration by parts to obtain:
\begin{align}
 \delta S_{\text{s}}=& \int d^4x\;\left[ \left(  {\cal E}_{\phi}'+ \left({\cal E}_{\psi} -{\cal E}_{\phi} \right)\frac{a'}{a}+ {\cal E}_{B}\frac{}{} -{\cal E}_{\varphi_1}\varphi'_0\right)\xi^{0}\right. \nonumber\\
&+ \left.\left( {\cal E}_{B}' - {\cal E}_{E}  \right)\xi \frac{}{}\right].
\end{align}

\noindent Since the action is gauge invariant and $\xi$ and $\xi^0$ are completely arbitrary, $\delta S_{\text{s}}=0$, and then both parenthesis are zero. This gives us two equations, which correspond to the Noether identities:
\begin{align}
{\cal E}_{\phi}'+ \left({\cal E}_{\psi} -{\cal E}_{\phi} \right)\frac{a'}{a}+  {\cal E}_{B} -{\cal E}_{\varphi_1}\varphi'_0&=0, \label{rel1} \\
{\cal E}_{B}'- {\cal E}_{E}&=0.\label{rel2}
\end{align}

\noindent Any field with an equation of motion that can be worked out from these last two identities, has a redundant equation of motion. Thus, we could use the gauge freedoms to set to zero in the action:
\begin{equation}\label{chos}
 (\psi,\varphi_1) + (E).
\end{equation}

\noindent Equation (\ref{chos})\footnote{\label{f1}This equation does not include $B$ as a possible field to be fixed by the gauge choice. Its equation of motion can, in fact, be worked out from (\ref{rel1}) but the problem is that $B$ cannot be fixed along with $E$. This is because the information in the equation for $E$ is contained in the equation for $B$ (see eq.~(\ref{rel2})). Thus, if $E$ is eliminated from the action we must keep $B$ to not lose information.} means that we can use one scalar gauge freedom to fix the value of one field from the first parenthesis and the other gauge freedom to fix one field from the second parenthesis. 

We must choose a gauge choice in such a way that the final fields are gauge-invariant in order to represent physical degrees of freedom. Our gauge choice will be:
\begin{equation}\label{gauge}
\varphi_1=0 \quad \mbox{and} \quad E=0.
\end{equation}

\noindent We have now reduced our initial problem with 5 scalar perturbation fields to one with 3: $\psi$, $B$ and $\phi$. Furthermore, from the second-order action we can observe that $B$ and $\phi$ have no dynamics; they are auxiliary fields that can be eliminated from the action by writing them in terms of $\psi$. If this is done, we finally obtain an action for $\psi$ only (in Fourier space):
\begin{equation}\label{SOApsi}
S_\text{s}[\psi]=\frac{1}{2} \int d^3k d\eta \frac{a^2 \varphi_0^{'2}}{\mathcal{H}^2} \left[ \psi^{'2} - k^2 \psi^2 \right].
\end{equation}

We can now compare our approach to the usual, gauge-invariant approach. There, the
variable of choice is the {\it comoving curvature perturbation}:
\begin{equation}
 \mathcal{R} \equiv \psi +\frac{\mathcal{H}}{\varphi'_0}\varphi_1,
\end{equation}

\noindent which is gauge invariant and turns out to be identical to the perturbative metric variable $\psi$ with our gauge choice:
\begin{equation}
 \mathcal{R}=\psi \quad \mbox{if} \quad \varphi_1=0.
\end{equation}
Unsurprisingly, we find that (\ref{SOApsi}) is the action for the only physical field in inflation, and this result is gauge-independent. 

The procedure we have described is systematic and straightforward and can easily be deployed to more complex theories. We now apply it to the EiBI theory for scalar and tensor perturbations. We neglect vector perturbations because, in our model, they decay as the universe expands, thus they are cosmologically irrelevant.

\section{The scalar second-order action for the E\lowercase{i}BI theory}
\label{MethodEiBI1} 
We can now apply our technique to bigravity, in particular the EiBI theory. In order to do that, we will initially consider general first-order perturbations and calculate a second-order action for {\it all} the perturbation fields present.  We will then reduce the obtained action to one containing only the physical degrees of freedom.

Let us start by considering first-order scalar perturbations for both metrics $q$ and $g$ on the background shown in Fig.~\ref{fig1}:
\begin{align}
 ds_q^2 &= b^2[z^{-1}(1+2\phi_1)d\eta^2-2B_{1,i}\sqrt{z}^{-1}dx^id\eta\nonumber\\
&-[(1-2\psi_1)\delta_{il}+2E_{1,il}]dx^i dx^l], \nonumber \\
 ds_g^2 &= a^2[(1+2\phi_2)d\eta^2-2 B_{2,i}dx^id\eta\nonumber\\
&-[(1-2\psi_2)\delta_{il}+2 E_{2,il}]dx^i dx^l]. \label{spert}
\end{align}
Here, the gravitational scalar fluctuations are described by eight functions $\phi_j$, $B_j$, $\psi_j$ and $E_j$, with $j=1,2$. For the matter part we will proceed as in \cite{Mukhanov:1990me} and consider a perfect fluid with an equation of state $p=w \rho$ and whose scalar fluctuation is described by one function $\chi(\eta,\vec{k})$. Consequently, there are 9 scalar perturbation fields in total. Since we are studying primordial perturbations, we will assume $w=1/3$, i.e.~we are in the radiation-dominated era. 

The next step is straightforward but long; we replace (\ref{spert}) and $\chi$ in action (\ref{EBI}) and Taylor expand up to second order, to find the second-order action $S_\text{s}[\phi_j, B_j, \psi_j, E_j, \chi]$: 
\begin{equation}\label{EBI2}
 S_\text{s}[\phi_j, B_j, \psi_j, E_j, \chi]= S_{1\text{s}}+S_{2\text{s}}+S_{\text{ms}},
\end{equation}
where, $S_{1\text{s}}$ is the first integral of (\ref{EBI}), i.e.~the gravitational action for $q$ purely; $S_{2\text{s}}$ is the second integral of (\ref{EBI}), i.e.~the coupling term between $q$ and $g$ plus a term for only $g$; and $S_{\text{ms}}$ is the second-order action of $S_\text{m}$ (in this case, the action of a perfect fluid). Explicitly, these three actions are the following:
\begin{align}\label{S1}
 &S_{1\text{s}}[\phi_1,B_1,\psi_1,E_1]=\frac{1}{2}\int d^4x\;  \frac{b^2}{\sqrt{z}} \left[4zh\psi_1' E_{1,ii}-6z\psi_1^{'2} \right.\nonumber\\
&-12zh(\phi_1+\psi_1)\psi'_1-2\psi_{1,i}(2\phi_{1,i}-\psi_{1,i} )-4h\psi_{1,i}B_{1,i} \nonumber \\
&+6zh^2(\phi_1+\psi_1)E_{1,ii}-4\sqrt{z}h(\phi_1+\psi_1)(B_1-\sqrt{z}E_1')_{,ii} \nonumber\\
&  -4\sqrt{z}\psi_1'(B_1-\sqrt{z}E_1')_{,ii}-4\sqrt{z}h E_{1,ii}(B_1-\sqrt{z} E'_1)_{,ll}    \nonumber \\
& +4\sqrt{z}hE_{1,ii}B_{1,ll}+3zh^2E_{1,ii}E_{1,ll} +3zh^2B_{1,i}B_{1,i}   \nonumber\\
&\left.-9zh^2(\phi_1+\psi_1)^2\frac{}{}\right]-\frac{b^4}{\kappa\sqrt{z}}\left[\frac{3}{2}\psi_1^2-3\phi_1\psi_1+\frac{1}{2}B_{1,i}B_{1,i}\right.\nonumber\\
& \left.-\frac{1}{2}E_{1,ii}E_{1,ll}-\frac{1}{2}\phi_1^2+E_{1,ii}(\phi_1-\psi_1)\right],
\end{align}
\begin{align}\label{S2}
&S_{2\text{s}}[\phi_j,B_j,\psi_j,E_j]=\frac{1}{2}\int d^4x \;\frac{a^2 b^2}{\kappa\sqrt{z}} \left[2\sqrt{z}B_{1,i}B_{2,i} \right. \nonumber\\ 
&  + \phi_1\left(\left(z-1\right)\left(3\psi_1-E_{1,ii}\right)-6\psi_2+2E_{2,ii}-2z\phi_2\right) \nonumber \\
& +\psi_1(6\psi_2-(z-1)E_{1,ii}-2E_{2,ii}-6z\phi_2)  \nonumber \\
&  -\frac{1}{2}(z-1)(E_{1,ii}E_{1,ll}+B_{1,i}B_{1,i}) +\frac{3}{2}\left(\phi_1^2+\psi_1^2\right)(z-1) \nonumber\\
&\left.-2E_{1,ii}\left(\psi_2-z\phi_2+E_{2,ii}\right)\right] -2a^4\left(\frac{3}{2}\psi^2_2-\frac{1}{2}\phi_2^2 \right.\nonumber \\
& \left. +\frac{1}{2}B_{2,i}B_{2,i}+(\phi_2-\psi_2)E_{2,ii} -3\phi_2\psi_2\right), 
\end{align}
\begin{align}\label{Sm}
&S_{\text{ms}}[\phi_2,B_2,\psi_2,E_2,\chi]=\int d^4x\; a^4\left[\frac{1}{2}\rho_0\phi_2^2+p_0\left(\frac{3}{2}\psi_2^2\right.\right. \nonumber\\
& \left.+(\phi_2-\psi_2) E_{2,ii}+\frac{1}{2}B_{2,i}B_{2,i}-\frac{1}{2}E_{2,ii}E_{2,ll}-3\phi_2\psi_2 \right)\nonumber\\
&-\frac{1}{6}(\rho_0+p_0)(3\psi_2-E_{2,ii}-\chi'_{,ii})^2 \nonumber\\
& \left. +(\rho_0+p_0)\left(\frac{1}{2}\chi_{,i}^{'2}+B_{2,i}\chi'_{,i}+\phi_2\chi_{,ii}\right) \right].
\end{align}
Here, we have used $h= b'/b$, $\rho_0$ is the background rest energy density of the fluid, and $p_0$ its pressure. 

With the full action in hand, we must now proceed to reduce it to one containing only physical perturbation fields by eliminating auxiliary variables and fixing the gauge. To do so we now study the gauge invariance of (\ref{EBI2}); we  apply the ideas described in the previous section, and look for the Noether identities for action (\ref{EBI2}). 

Our full action is invariant under the following infinitesimal transformations:
 \begin{align}\label{transf}
 & \delta \phi_2=-\frac{a'}{a}\xi^0-\xi^{0'},\;\delta B_2=\xi^0-\xi' \; \delta\psi_2=\frac{a'}{a}\xi^0,\nonumber \\
 & \delta E_2=-\xi, \; \delta\phi_1=-\left[\frac{b'}{b}-\frac{z'}{2z}\right]\xi^0-\xi^{0'},\; \delta\chi=\xi, \nonumber \\
&  \delta B_1=-\xi'\sqrt{z}+\frac{\xi^0}{\sqrt{z}},\; \delta\psi_1=\frac{b'}{b}\xi^0,\; \delta E_1=-\xi, 
\end{align}
where $\xi$ and $\xi^0$ are infinitesimal arbitrary functions that represent the two gauge freedoms in the case of scalar perturbations. This leads us to two Noether identities (one for each gauge freedom)
\begin{align}\label{NoethEBI}
   &{\cal E}_{\phi_1}'-{\cal E}_{\phi_1}\left[ \frac{b'}{b}-\frac{z'}{2z}\right]+ {\cal E}_{\psi_1}\frac{b'}{b}+\frac{{\cal E}_{B_1}}{\sqrt{z}} +{\cal E}_{\phi_2}'\nonumber \\
&+ \left({\cal E}_{\psi_2} -{\cal E}_{\phi_2} \right)\frac{a'}{a}+  {\cal E}_{B_2} = 0, \nonumber\\
 & ({\cal E}_{B_1}\sqrt{z})' - {\cal E}_{E_1}- {\cal E}_{E_2}+ {\cal E}_{B_2}'+ {\cal E}_{\chi} =0.
\end{align}
which we can use to fix the value of the following fields:
\begin{equation}\label{choice}
 (\psi_1, \psi_2) + (E_1, E_2, \chi).
\end{equation}
We will choose:  
\begin{equation}\label{ch1}
\psi_1=0 \quad \mbox{and} \quad \chi=0.
\end{equation} 
Analogously to the previous section, the gauge-invariant {\it curvature perturbation on slices of uniform energy density}:
\begin{equation}\label{Invariant}
\zeta \equiv \psi_2-\frac{1}{3(\rho_0+p_0)}\delta \rho,
\end{equation}
where $\delta \rho$ is the first-order energy density fluctuation\footnote{$\delta \rho$ is given by $\delta T^0{}_0$, which in terms of the perturbation fields is given by $\delta \rho=(\rho_0+p_0)(3\psi_2-E_{2,ii}-\chi_{,ii})$.}, turns out to be proportional to the perturbative metric variable\footnote{In principle, $\zeta$ is proportional to $E_2$, but as we will see later, $E_2=E_1$.} $E_1$ in Fourier space in our gauge choice (\ref{ch1}):
\begin{equation}\label{relEZ}
\zeta = -\frac{1}{3}k^2 E_1 \quad \mbox{if} \quad \chi=0,
\end{equation}
where $k$ is the comoving wavenumber of the perturbation. This variable $\zeta$ is sometimes used to describe the physical scalar field in the inflation theory, as an alternative to $\mathcal{R}$. As we will see later, there is only one physical scalar field in this theory, and it can be described by $E_1$.

We now proceed with the reduction process of action (\ref{EBI2}). For simplicity, we will work in Fourier space. First of all, we fix the gauge. Second,  we notice that the metric $g_{\mu\nu}$ enters with no derivatives in the action (\ref{EBI}). This means that the fluctuations $\phi_2$, $B_2$, $\psi_2$ and $E_2$ are auxiliary variables. Furthermore, analogously to inflation, we can see from (\ref{S1})-(\ref{S2}) that $\phi_1$ and $B_1$ are also auxiliary variables. Thus, two functions are fixed by gauge conditions, and other six function can be eliminated, resulting in only one physical field present, $E_1$. 

We now use our gauge choice (\ref{ch1}) in Fourier space, and consider the mode expansion of the remaining fields:  
\begin{eqnarray}
 &\phi_j(\eta,\vec{x})=\int\frac{d^3k}{(2\pi)^{\frac{3}{2}}}\phi_j(\eta,\vec{k})e^{i\vec{k}\cdot\vec{x}},\nonumber\\
& B_j(\eta,\vec{x})=\int \frac{d^3k}{(2\pi)^{\frac{3}{2}}}B_j(\eta,\vec{k})e^{i\vec{k}\cdot\vec{x}}, \nonumber\\
& \psi_2(\eta,\vec{x})=\int \frac{d^3k}{(2\pi)^{\frac{3}{2}}}\psi_2(\eta,\vec{k})e^{i\vec{k}\cdot\vec{x}}, \nonumber \\
& E_j(\eta,\vec{x})=\int \frac{d^3k}{(2\pi)^{\frac{3}{2}}}E_j(\eta,\vec{k})e^{i\vec{k}\cdot\vec{x}},
\end{eqnarray}
where $j=1,2$, and $k^2\equiv \vec{k}\cdot\vec{k}$. Note that $\chi=\psi_1=0$, in our gauge choice. Varying the whole action $S_\text{s}$ with respect to these scalar perturbation fields in Fourier space, we obtain the following equations of motion:
\begin{align}\label{eqnsg}
 \phi_2 :\;& \left(\phi_1-\phi_2+3\psi_1+E_1k^2\right)z -E_2k^2 -3\psi_2=0,\nonumber\\
 \psi_2 :\;&   3(\phi_2-\phi_1)-\left(E_2k^2+3\psi_2\right)z + E_1k^2 =0,\nonumber \\
 E_2 :\;&   3(\phi_1-\phi_2) +k^2\left( 3E_1-4E_2\right)+z\left( 3\psi_2+k^2 E_2\right) =0,\nonumber\\
 B_2 :\;&  B_2-\sqrt{z}B_1=0,
\end{align}
for the metric $g$, and:
\begin{align}
\phi_1 :\;& a^2 z\phi_2+3a^2\psi_2 -2k^2\left(\kappa h\sqrt{z} B_1+\frac{1}{2} a^2\left(E_1-E_2\right)\right)\label{eqnphi1} \nonumber\\
&+2 z \kappa h E_1' k^2-\left(3 a^2-2b^2\right)\phi_1 =0, \\
 B_1 :\;&  a^2 \sqrt{z} B_2- B_1a^2+2 \kappa \sqrt{z}h\phi_1=0, \label{eqnB1}\\
 E_1 :\;& a^2\left(\psi_2+k^2 \left(E_1-E_2\right)-z\phi_2\right) + \left(2b^2-a^2\right)\phi_1 \label{eqnE1}\nonumber\\ 
& + 2\kappa z h \phi_1' =0,
\end{align}
for the metric $q$.
From the set of equations (\ref{eqnsg}) we can obtain all the perturbation fields for $g$: $E_2$, $B_2$, $\phi_2$, and $\psi_2$, in terms of the perturbation fields of $q$:
\begin{align*}
 &  \phi_2=  \frac{(3+z^2)\phi_1+k^2E_1(z+1)(z-1)}{3+z^2},\\  
&  \psi_2= \frac{-k^2(z-1)(1/3z-1)E_1}{3+z^2}, \\
& B_2= \sqrt{z}B_1, \\
&E_2=E_1.
\end{align*}
If we replace these results in the remaining three equations (\ref{eqnphi1})-(\ref{eqnE1}), we can see that $\phi_1$ and $B_1$ can be expressed in terms of $E_1$,
\begin{align}
 \phi_1&= -\frac{\left(z-1\right)\left(\kappa zh E_1'+\frac{1}{2} E_1 \left(z-1\right) a^2\right)k^2a^2}{\left(3\left(z-1\right)a^2+2\kappa k^2\right)\kappa h^2 z},\\
 B_1 &= \frac{2\kappa h zE_1' k^2+ k^2 a^2 E_1\left(z-1\right)}{\left(3 a^2\left(z-1\right)+2\kappa k^2\right)h\sqrt{z}}.
\end{align}

Finally, we have obtained all fields in terms of $E_1$. If we write these fields in terms of $\zeta$, by using the relation (\ref{relEZ}), and replace them in action (\ref{EBI2}), we get the final reduced action $S_{\text{s}}[\zeta]$ in Fourier space: 
\begin{equation}\label{Szeta}
S_{\text{s}}[\zeta] = {1 \over 2}\int  d^3k d\eta\; f_1(\eta,k)  \left(  \zeta^{'2} - f_2(\eta,k)  \zeta^2 \right)  
\end{equation}    
where $f_1$ with $f_2$ are time-dependent functions of the background variables given by:
\begin{align}
&f_1(\eta,k) = {18 b^2 a^2 (z-1) \sqrt{z} \over X}\nonumber \\
&f_2(\eta,k) =  9b^2 a^2  \left\{ 2\mathcal{H}\left(z^2+3\right)\kappa \left[ 9\left(z-2\right)\left(z-1\right)^2 a^4 \right.\right.\nonumber \\
& \left. +12k^2\kappa \left(z-1\right)\left(z-2\right)a^2 + 2k^4 \kappa^2\left(z-3\right) \right] \nonumber \\
&  + z'\left(z^2+3\right)\left[9\left(z-1\right)^2 a^4 + 12 k^2\kappa \left(z-1\right) a^2 + 2\kappa^2 k^4\right] \nonumber\\
& \left.+ hX\kappa \left[12za^2(z-1)(1+z^2)+ (3+5z+z^2+7z^3)k^2\kappa\right] \right\}  \nonumber\\
&/ \left\{X^2 \kappa^2 h\sqrt{z}\left(z^2+3\right)f_1\right\},
\end{align}   
where $X\equiv 3\left(z-1\right)a^2 + 2k^2\kappa$. 
The result given in (\ref{Szeta}) is gauge-independent and describes the scalar perturbations in the EiBI theory.

\section{The tensor second-order action for the E\lowercase{i}BI theory}
\label{MethodEiBI2}

Let us now consider first-order tensor perturbations:
\begin{align}
 ds_q^2&=b^2[z^{-1}d\eta^2-(\delta_{ij}+h_{1il})dx^i dx^l],\nonumber \\
ds_g^2&=a^2[d\eta^2-(\delta_{ij}+h_{2il})dx^i dx^l],
\end{align}
where the tensor fluctuations are given by $h_{1il}$ and $h_{2il}$. We can identify two  polarisations $p=(+,\times)$ and, for simplicity, we will choose a specific direction $\vec{k}=k\hat{z}$ so tensor perturbations lie in the $xy$ plane. As a result, equivalently, tensor metric perturbations can be written as:
\begin{align}
ds^2_q =&b^2\left[z^{-1}d\eta^2-[(1+h_{1+})dx^2+(1-h_{1+})dy^2\nonumber \right.\\
&\left. + dz^2+2h_{1\times} dxdy]\right]\nonumber \\
ds^2_g =&a^2\left[d\eta^2-[(1+h_{2+})dx^2+(1-h_{2+})dy^2\right.\nonumber\\
&\left. +dz^2+2h_{2\times} dxdy]\right]
\end{align}
where the perturbation fields are $h_{jp}(z,\eta)$, for $j=1,2$. Replacing these expressions into (\ref{EBI}) we obtain a second-order action $S_\text{T}[h_{jp}]$:
\begin{equation}\label{tensorS}
S_\text{T}[h_{jp}]= S_\times[h_{j\times}] + S_+[h_{j+}],
\end{equation}

\noindent where
\begin{align}\label{Sp1}
&S_{p}[h_{jp}]= \frac{1}{2}\int d^4x\; \frac{b^2}{2\sqrt{z}}\left[ zh_{1p}^{'2}+ \frac{2}{\kappa}a^2h_{1p}^2 - h_{1p,z}^2\right.\nonumber\\
&\left.+\frac{2}{\kappa}a^2 \left(h_{2p}^2-2h_{1p}h_{2p}\right)\right].
\end{align}
As in the standard GR case, there is no matter contribution to tensor perturbations. 

The action in expression (\ref{tensorS}) does not have a gauge symmetry, so all we need to do is eliminate the auxiliary variables. We take the mode expansion,
\begin{equation}
h_{jp}(\eta,\vec{x})=\int \frac{d^3k}{(2\pi)^{\frac{3}{2}}}h_{jp}(\eta,\vec{k})e^{i\vec{k}\cdot\vec{x}}, 
\end{equation}
and calculate the equations of motion,
\begin{eqnarray}
h_{2p}:\; & h_{2p}-h_{1p}=0,\label{eqnh2}\\
  h_{1p}:\; & h_{1p}^{''}+ \left(2h+\frac{z'}{2z}\right)h_{1p}' +\left(\frac{k^2}{z}+\frac{2}{\kappa z}a^2\right)h_{1p}\label{eqnh1}\nonumber \\
&-\frac{2}{\kappa z}a^2h_{2p}=0.
\end{eqnarray}
From equation (\ref{Sp1}) we can see that only $h_{2p}$ are auxiliary variables. Using eq.~(\ref{eqnh2}) we can work out $h_{2p}$ in terms of $h_{1p}$ and replace them in action (\ref{Sp1}) to obtain the reduced second-order action in Fourier space:
\begin{equation}
S_\text{T}[h_{1p}]= S_\times[h_{1\times}] + S_+[h_{1+}],
\end{equation}
where
\begin{equation}\label{Sp}
S_p[h_{1p}]=\frac{1}{2}\int  d^3k d\eta\; b^2\sqrt{z}\left( h_{1p}^{'2}-\frac{k^2}{z}h_{1p}^2\right).
\end{equation} 
We see that $S_\text{T}[h_{1p}]$ has two copies of the same action $S_p[h_{1p}]$ for each polarisation; from now on we will describe these two physical degrees of freedom as $h$, such that $h=h_{1\times}=h_{1+}$.

\section{Quantisation and Cosmological Predictions}
\label{Quantum}

We have reached the final step of the calculation. We can now proceed to the quantisation of the physical fields $\zeta$ and $h$. But, before we do so, we would like to comment on a particular behaviour of the classical solutions during the Eddington regime.  The equation of motion coming from (\ref{Szeta}) for $\zeta$ can be approximated to first order in the Eddington regime ($a\approx a_B$):
\begin{equation}\label{sole1}
 \zeta^{''}\approx 0  \quad \Rightarrow \quad  \zeta(\eta,k)=A_k\eta+B_k, 
\end{equation}
where $A_k$ and $B_k$ are integration constants, depending on $k$. From this solution we can see that there is a linear divergence as $\eta\rightarrow -\infty$ (or $a\rightarrow a_B$), leading to a scalar instability. Even though, in this paper, we will not consider the case $\kappa<0$, we would like to mention that there too, there is a scalar polynomial instability as $a\rightarrow a_B$. The same behaviour was found in \cite{EscamillaRivera:2012vz} for the tensor perturbation $h$ in both cases of $\kappa$.

Since the amplitude of these fields was large near $a_B$, at some moment in the past, the linear theory of perturbations breaks down because $\delta g_{\mu\nu}/ g_{\mu\nu}^{(0)} \ll 1$ is violated (where $g_{\mu\nu}^{(0)}$ is the background metric and $\delta g_{\mu\nu}$ its perturbation). This problem can be caused by the linear perturbation theory (corrections of higher order could change this behaviour) or could be a characteristic of the EiBI theory itself. A similar problem appears in inflation, where the physical fields diverge in the big bang. However, we will consider only the region where the linear theory is still valid, and thus avoid this problem. This means we cannot fully exploit one of the attractive features of the EiBI theory: that it will have existed for an arbitrarily long amount of time in the Eddington regime before the Einstein regime began.

To quantise the perturbations in this model, we apply the canonical quantisation procedure by promoting the fields to quantum operators, expanding in terms of annihilation and creation operators and imposing commutation relations. These quantum solutions are not completely determined; one initial condition is missing, which leaves the vacuum state undefined. In the inflation theory of GR it is assumed that perturbations were originated in their ground state- the Bunch-Davies vacuum- at sub-Hubble scales (the period when a comoving wavelength $\lambda$ is much smaller than the comoving Hubble radius), which determines entirely the quantum solutions (\cite{Mukhanov:2005}). It is possible, however to choose different vacua (or, equivalently, different initial conditions) for perturbations  (see \cite{Allen:1985ux,Greene:2005wk,Chung:2003wn,Danielsson:2002kx}), which lead to different results for the quantum solutions. In principle a complete theory of quantum gravity could tell us exactly how the universe leaves the Planck scale, and would give us a unique prescription for the initial condition for the fluctuations. 

An alternative proposal is given by Hollands and Wald in \cite{Hollands:2002yb}: a mechanism that results in a scale-invariant power spectrum for quantum fluctuations in GR, without assuming the existence of a fundamental scalar field as matter. To do so, let us assume that there is a fundamental length called $l_0$. Semiclassical physics applies to phenomena on spatial scales larger that $l_0$, so modes emerge from an unknown fundamental description of spacetime at that scale. We may, for example, assume that a perturbation with physical wavelength $\lambda_{ph}$ is effectively born at $l_0$ in the ground state of a flat spacetime. Since $\lambda_{ph}$ grows in time, the perturbations are continuously being created. When applied to scalar perturbations in GR during the early universe, Hollands and Wald obtained a primordial exactly scale-invariant power spectrum, almost in accordance with observations (\cite{Ade:2013uln}). In order to have the correct amplitude for the power spectrum, the authors choose $l_0$ to be $l_0=10^{5}l_p$, with $l_p$ the Planck scale.

We can apply the Hollands-Wald mechanism to the quantum solutions $\hat{\zeta}(\eta,\vec{k})$ and $\hat{h}(\eta,\vec{k})$, albeit with a slight modification. We will define\footnote{We have chosen this order of magnitude of $l_0$ in order to have the correct order of magnitude for the power spectrum of scalar perturbations.} $l_0=\sqrt{|\kappa|}= 10^4 l_p$, and assume that a mode with comoving wavenumber $k$ is created at $\eta_*$ such that $b_*/k= l_0$. Notice that we have defined this relation\footnote{This modification is motivated by the form of the second-order action for tensor perturbations. In GR the action for $h$ has the form of an action for a scalar field coupled to the metric $g$, but in the EiBI theory it appears coupled to $q$. } with the scale factor of the metric $q$, instead of $g$. Then, the initial condition for the perturbations will be that they are in the ground state at $\eta_*$. For all cosmologically relevant scales, $\eta_*$ occurs in the Eddington regime.

Since we will use the Hollands-Wald mechanism with $q$ as the main metric, we will rewrite (\ref{Szeta}) such that it looks like an action for a scalar field in a background described by $q$:
\begin{equation}
 S_{\text s}[v]=\frac{1}{2}\int d^3k d\eta \; b^2\sqrt{z}\left(v^{'2}-f_3(\eta,k) v^2\right),
\end{equation}
where the field $v$ is related to $\zeta$ through $\zeta=v\sqrt{\frac{b^2\sqrt{z}}{f_1}}$, and $f_3$ is a function depending on $f_1$ and $f_2$.
Also, since $\eta_*$ occurs in the Eddington regime, we make the approximation $a\approx a_B$ and find
\begin{equation}\label{Actionvs}
S_\text{s}\approx \frac{1}{2}\int d^3k d\eta \; 4a_B^2\left( v^{'2}-\frac{5k^2}{9a_B}(a-a_B)v^2 \right).
\end{equation}
We write the quantum solution of $v$ as:
\begin{equation}
\hat{v}(\eta,\vec{k})= v_{\vec{k}}a_{\vec{k}} + v^*_{\vec{k}}a_{\vec{k}}^\dagger,
\end{equation}
where $v_{\vec{k}}$ is a classical complex solution, and $a_{\vec{k}}$ and $a_{\vec{k}}^\dagger$ are the annihilation and creation operators, respectively. By making an adiabatic approximation\footnote{The adiabatic approximation consists in taking an interval of time small enough to allow us to consider the background functions as effectively constant. In our specific calculations, we take an interval of time around $\eta_*$.} near $\eta_*$ in the action (\ref{Actionvs}), we can use the standard QFT rules (with $\hbar=1$) to quantise $v$ and then write $v_{\vec{k}}$ at $\eta_*$ as:
\begin{equation}
v_{\vec{k}}(\eta_*)=\frac{1}{\sqrt{8a_B^2\omega_*}}e^{i\omega_* \eta_*};\quad \omega_*=\sqrt{\frac{5k^2}{9a_B }(a_*-a_B)},
\end{equation}
which gives us the initial condition for $\zeta$:
\begin{equation}\label{IniCondZ}
 \zeta_{\vec{k}}(\eta_*)=-\sqrt{\frac{1 }{4\sqrt{5}\kappa k^3}}e^{i\omega_* \eta_*}, 
\end{equation}
such that 
\begin{equation}
 \omega_*=\frac{\sqrt{5}k^3\kappa}{12a_B^2}; \quad \eta_*=\frac{2\sqrt{3\kappa}}{\sqrt{2}a_B}\ln\left( \frac{k\sqrt{\kappa}}{2a_B}\right) .
\end{equation}
Here, we have used that $l_0=b_*/k$.

Analogously, we approximate the tensor action for $h$ in the Eddington regime:
\begin{equation}
S_{p}\approx \frac{1}{2}\int d^3k d\eta\; 4a_B^2\left(h^{'2} -\frac{k^2}{a_B}(a-a_B)h^2\right),
\end{equation}
 
\noindent and write the quantum solution as:
\begin{equation}
\hat{h}(\eta,\vec{k})= h_{\vec{k}}a_{\vec{k}} + h^*_{\vec{k}}a_{\vec{k}}^\dagger,
\end{equation}
where $h_{\vec{k}}$ is a complex classical solution. Taking the adiabatic approximation we find the following initial condition for $h_{\vec{k}}$ at $\eta_*$:
\begin{equation}\label{IniCondh}
h_{\vec{k}}(\eta_*)=\sqrt{\frac{1}{2\kappa k^3}}e^{i\tilde{\omega}_* \eta_*},
\end{equation}
such that
\begin{equation}
\tilde{\omega}_*=\frac{\kappa k^3}{4a_B^2}; \quad \eta_*=\frac{2\sqrt{3\kappa}}{\sqrt{2}a_B}\ln\left( \frac{k\sqrt{\kappa}}{2a_B}\right)
\end{equation}
We now have the initial conditions for both quantum scalar and tensor perturbations. A general solution can be obtained by evolving in time the classical solutions $\zeta_{\vec{k}}$ and $h_{\vec{k}}$ with their equations of motion. However, this extrapolation seems difficult analytically, so we do it numerically.

We now want to compare with observations. Current constraints pin down properties of the  scalar and tensor power spectra at the time when a given scale leaves the horizon (which is the second time a perturbation crosses the horizon), i.e.~$k=\mathcal{H}$ which coincides with the transition from quantum perturbations to classical perturbations.  We now proceed to predict these power spectra.

The power spectrum of $\zeta$ is defined as:
\begin{equation}
\mathcal{P}_{\zeta}(\eta,k)=\frac{k^3}{ 2\pi^2} |\zeta_{\vec{k}}(\eta)|^2.
\end{equation}
In order to find $\mathcal{P}_{\zeta}$  we will extrapolate $\zeta_{\vec{k}}$ numerically in time by using the classical equation of motion and the initial condition (\ref{IniCondZ}) at $\eta_*$. This initial condition is imposed on $\zeta$ and its derivative. We evaluate the numerical solution at a particular time during the Einstein regime for super-Hubble scales (while the perturbation wavelength is larger than the horizon), and calculate $|\zeta_{\vec{k}}|^2$ there for a range of values of $k$.  We find that that the power spectrum is scale-invariant: here is no dependence on the value of $k$, i.e.
\begin{equation}
\mathcal{P}_{\zeta}(k)=A_\zeta^2 k^{n_{\text s}-1},
\end{equation}
where the amplitude $A_\zeta^2\sim 10^{-9}$, and the scalar spectral index is perfectly scale invariant, $n_{\text s}-1 =0$.
 

To calculate the power spectrum of $h$ we proceed as for the scalar field $\zeta$. We perform numerical calculations to find $|h_{\vec{k}}|^2$ and conclude that the power spectrum is nearly scale-invariant:
\begin{equation}
\mathcal{P}_{\text T}(k)= \frac{2 k^3}{\pi^2} |h_{\vec{k}}|^2 = A_{\text{T}}^2k^{n_{\text T}},
\end{equation}
where the amplitude $A_{\text T}^2\sim 10^{-8}$ the tensor spectral index is also perfectly scale invariant $n_{\text T}=0$. 

We can now compare our predictions for the power spectra of scalar and tensor perturbations with the results obtained from observations\footnote{Usually experimental results refer to the comoving curvature perturbation $\mathcal{R}$, instead of $\zeta$. However, for super-Hubble scales $\mathcal{R}\approx \zeta$.}. Using
\begin{equation}
\mathcal{P}_\mathcal{\zeta}(k)=A_{\mathcal{\zeta}}^2\left(\frac{k}{k_0}\right)^{n_\text{s}-1},
\end{equation}

\noindent where $k_0=0.05$ $\text{Mpc}^{-1}$ we have that \cite{Ade:2013uln}
\begin{align}
\mathcal{P}_\mathcal{R}(k_0)&=\left(2.196^{+ 0.051}_{- 0.060}\right)\times 10^{-9},\nonumber\\
 n_{\text s}-1 &=-0.0397\pm 0.0073.
\end{align} 
We found that the amplitude of the scalar power spectrum has the right order of magnitude (the exact value can be fitted by choosing an appropriate value for $\kappa$), while the predicted $n_{\text{s}}$ is not compatible with current constraints. The maximum value for the tensor-to-scalar ratio $r(k)$ has also been measured for $k_0$ (see \cite{Ade:2013uln}) as
\begin{equation}
r(k_0)\equiv \frac{\mathcal{P}_\text{T}(k_0)}{\mathcal{P}_\mathcal{R}(k_0)} < 0.11,
\end{equation}
We predict $r\sim 10$ which means that the EiBI theory with the Hollands-Wald prescription (in particular, with the described choice of vacuum) is grossly inconsistent with observation. 
\section{Conclusions}
 \label{Conclusion}
In this paper we have proposed a general algorithm for studying linear cosmological perturbations in multi-gravity. We have applied it to explore the possibility that inhomogeneities in our universe  were generated by primordial quantum first-order perturbations in the classical cosmological background given by the EiBI theory.  By carefully exploring the gauge symmetry present in the theory, we were able to write an action which only contains the physical degrees of freedom: one scalar and two equal tensor perturbations. The beauty of our method is that it can be rolled out to other theories involving multiple metrics- the currently popular bigravity theories of massive gravity come to mind as well as more elaborate multigravity models. In some sense, the method we used is the core result of the paper. We have shown it to be powerful, unambiguous and straightforward. 

Since most of multigravity theories are rather new, at present the cosmological studies are focused on the evolution of a homogeneous and isotropic universe matching the observations (\cite{Bamba:2013hza,Comelli:2011zm,Khosravi:2011zi,vonStrauss:2011mq,Akrami:2012vf}). The application of the method developed in this paper to these theories should be very similar and could help to go further and study the evolution of primordial quantum perturbations, giving more observational constraints.

We then used the canonical formalism to quantise the scalar and tensor physical fields. We argued that there were ambiguities in making a vacuum choice and opted for the Hollands-Wald mechanism. With this procedure, we found perfectly scale-invariant power spectra for scalar and tensor perturbations with a large tensor-to-scalar ratio, grossly inconsistent with current observations.  This means that, in its current incarnation, the EiBI theory is not a viable model for the early universe. This is not the end of the road and there are unexplored avenues. For a start we have considered the minimal model, with no extra fields. If one were to embrace the presence of other states of equation or fundamental fields at early times, it may be possible to circumvent the problems that we found, as the solution for the early universe depends on the type of matter considered (see \cite{Scargill:2012kg} for some different cases). For instance, with the presence of a scalar field, at least the tensor instability can be avoided for $\kappa<0$, as described in \cite{Avelino:2012ue} and non-minimal couplings can further complicate the scenario (see \cite{Deser:1998rj}). Further research is needed to predict quantum perturbations in this scenario, but these preliminary findings already suggest an improvement of the minimal model. 

Furthermore, in working with the minimal model, we were forced to consider the Hollands-Wald mechanism for setting up the initial conditions in order to get viable results. Again, extra degrees of freedom may enlarge the space of possibilities for the quantum initial state leading to a more viable cosmology. All these possibilities merit further scrutiny for it would be truly intriguing if it were possible to have a viable cosmology that could emerge from a non-singular initial state.

\begin{acknowledgments}

\textit{Acknowledgments.---} We are grateful to Tessa Baker, Johannes Noller and James Scargill for discussions. PGF acknowledges support from Leverhulme, STFC, BIPAC and the Oxford Martin School and the hospitality of the Higgs Centre in Edinburgh. MB was partially supported by Fondecyt (Chile) \#1100282 and Anillo ACT (Chile) \#1102. ML was funded by Becas Chile.

\end{acknowledgments}

\appendix

\section*{Appendix: An example of a ``good" gauge choice.}
\label{App1}
Consider the action
\begin{equation}
S[A_0,\psi] = {1 \over 2}\int d^4x\; (A_0 - \dot\psi)^2, \label{maxwell}
\end{equation}
where $\dot\psi=d\psi/dt$. This action corresponds to the gauge part of Maxwell's theory, and it is invariant under the following gauge transformation:
\begin{equation}\label{MG}
\delta \psi =\tilde{\psi}-\psi= \epsilon(x), \quad  \delta A_0 = \tilde{A}_0-A_0= {d \over dt}\epsilon(x),
\end{equation}
where $\epsilon(x)$ is an arbitrary function, and $\tilde{\psi}$ and $\tilde{A}_0$ are the new fields. The existence of this gauge symmetry, means that there is one gauge freedom that could be used to fix the value of one field in the action (\ref{maxwell}), by choosing a particular value for the function $\epsilon(x)$. Consider the following two gauge choices:
\begin{itemize}
\item[1.] Set $\epsilon=\psi$: If we perform a gauge transformation with this $\epsilon$ in the action (\ref{maxwell}), then $\tilde{\psi}=0$, and the gauge-fixed action becomes 
\begin{equation}
S[\tilde{A}_0] = {1 \over 2}\int d^4x\; \tilde{A}_0^2,
\end{equation}
whose equation of motion is $\tilde{A}_0=0$, and therefore there are no degrees of freedom in this action. Here, the equation of motion for $\tilde{\psi}$ was lost.
\item[2.] Set $\dot\epsilon=A_0$: If we perform a gauge transformation with this $\epsilon$, then $\tilde{A}_0=0$ and the gauge-fixed action becomes 
\begin{equation}
S[\tilde{\psi}] = {1 \over 2}\int d^4x\; \dot{\tilde{\psi}}^2,
\end{equation}
whose equation of motion is $\ddot{\tilde{\psi}}=0$, and therefore there is one degree of freedom in this action. Here, the equation of motion for $\tilde{A}_0$ was lost.
\end{itemize}

Since the action is gauge invariant we expect to have the same result for any choice of $\epsilon$ at level of the equations of motion (fixing gauges in the equations of motion), but not at level of the action (fixing gauges in the action). In fact, this discrepancy at level of the action is noticed in our example where both gauge-fixed actions do not contain the same information; the first one does not describe any physical dynamic field, while the second does. The result is different because the second gauge choice is incorrect due to the fact that the equation of motion lost had crucial information that the remaining equation did not. In general, the tool to decide which choice is correct is the set of Noether identities. To obtain the Noether identities in our example, we consider a variation of the action (\ref{maxwell}):
\begin{equation}
\delta S=\int d^4x\;  \left({\cal E}_{A_0}\delta A_0+{\cal E}_{\psi}\delta \psi\right),
\end{equation}
where ${\cal E}_{A_0}$ and ${\cal E}_{\psi}$ are the equations of motion for $A_0$ and $\psi$, respectively. Now, we replace these variations by (\ref{MG}), obtaining:
\begin{equation}\label{GaugeSMax}
\delta S= \int d^4x\;\left({\cal E}_\psi -{d \over dt} {\cal E}_{A_0}\right)\epsilon,
\end{equation}
where some integration by parts were made. Since the action is gauge invariant and $\epsilon$ is arbitrary, the parenthesis of (\ref{GaugeSMax}) is zero:
\begin{equation}
{\cal E}_\psi = -{d \over dt} {\cal E}_{A_0},  \label{noether}
\end{equation}
representing the Noether identity\footnote{It is important to emphasize that Noether identities are universal, depending only on the gauge transformations and not on the particular action. In general, there will be one Noether identity for each gauge freedom present in a theory.} for this system.
Eq. (\ref{noether}) displays in a clear way the difference between both fields:
${\cal E}_{A_0}=0$ implies ${\cal E}_\psi=0$ while ${\cal E}_\psi=0$ does {\it not} imply ${\cal E}_{A_0}=0$. This is why it is correct to dispose of $\psi$, because its equation is already contained in the $A_0$ equation. The converse is not true and it is incorrect to fix the gauge with a condition on $A_0$. 

In general, it will be correct to use the gauge freedom to dispose of the fields that have a redundant equation of motion (i.e.~its information is contained on the remaining equations), what can be seen from the Noether identities.

 
\bibliographystyle{apsrev4-1}
\bibliography{Ref1}

\end{document}